\documentclass[journal=nalefd,manuscript=letter,layout=onecolumn]{achemso}
\usepackage[version=3]{mhchem} % Formula subscripts using \ce{}

\author{Xiaoyan Zhou}
\email{xiaoyan.zhou@nbi.ku.dk}
\author{Ravitej Uppu}
\author{Zhe Liu}
\author{Camille Papon}
\affiliation{Center for Hybrid Quantum Networks (Hy-Q), Niels Bohr Institute, University of Copenhagen, Blegdamsvej 17, 2100-DK Copenhagen, Denmark}

\author{R\"{u}diger Schott}
\author{Andreas D. Wieck}
\author{Arne Ludwig}
\affiliation{Lehrstuhl f\"{u}r Angewandte Festk\"{o}rperphysik, Ruhr-Universität Bochum, Universit\"{a}tsstrasse 150, D-44780 Bochum, Germany}

\author{Peter Lodahl}
\author{Leonardo Midolo}
\affiliation{Center for Hybrid Quantum Networks (Hy-Q), Niels Bohr Institute, University of Copenhagen, Blegdamsvej 17, 2100-DK Copenhagen, Denmark}
\title[NOEMS filter]
  {On-chip nanomechanical filtering of quantum-dot single-photon sources}


\begin{document}
\begin{abstract}
Semiconductor quantum dots in photonic integrated circuits enable scaling quantum-information processing to many single photons and quantum-optical gates. On-chip spectral filters are essential to achieve high-purity and coherent photon emission from quantum dots embedded in waveguides, without resorting to free-space optics. Such spectral filters should be tunable, to compensate for the inhomogeneous spectral distribution of the quantum dots transitions. Here, we report an on-chip filter monolithically integrated with quantum dots, that uses nanomechanical motion for tuning its resonant wavelength over 10 nm, enabling operation at cryogenic temperatures and avoiding cross-talk with the emitter. We demonstrate single-photon emission from a quantum dot under non-resonant excitation by employing only the on-chip filter. These results are key for the development of fully-integrated de-multiplexing, multi-path photon encoding schemes, and multi-emitter circuits. 
\end{abstract}

\section{}
Photonic integrated circuits offer a versatile platform for scaling quantum photonic technology to multi-qubit applications, reducing the experimental overhead and boosting the overall performance of quantum protocols \cite{o2009photonic,Carolan2015universal,flamini2018photonic}. The integration of quantum emitters in such circuits fundamentally changes the perspectives on chip-scale quantum information processing, as it enables deterministic emitter-photon\cite{arcari2014near} and spin-photon\cite{luxmoore2013interfacing,ding2019coherent} interfaces, coherent single-photon generation, \cite{wang2019towards,kirvsanske2017indistinguishable} and photon non-linearities \cite{javadi2015single}. Self-assembled indium aresenide (InAs) quantum dots (QDs) in gallium arsenide (GaAs) nanophotonic structures constitute an excellent platform for quantum photonics, combining well-characterized and outstanding emitter properties with a planar circuit technology\cite{lodahl2015interfacing,heppsemiconductor}. For example, on-chip photon routers\cite{papon2019nanomechanical} and detectors\cite{sprengers2011waveguide,reithmaier2015chip,schwartz2018fully} have been recently demonstrated in GaAs, suggesting a potential avenue for scaling GaAs-based quantum-photonic circuits towards complex applications within quantum communication or quantum simulation\cite{lodahl2017quantum}.
A key missing functionality to achieve multi-emitter or multi-photon integrated devices, is the on-chip spectral filtering of the emitter, required to 1) remove the signal from other QDs (or other excitonic states in the same QD) with non-resonant excitation schemes\cite{Santori2001triggered}, 2) improve the indistinguishability of the emitted photons by filtering phonon side-bands \cite{iles2017phonon}. In most experiments, filtering is performed off-chip, using etalons or gratings providing high extinction, low loss, and wavelength tunability\cite{kirvsanske2017indistinguishable,Dusanowski2019,wang2019towards}. The last requirement stems from the randomness of QD nucleation during sample growth, which results in a $\sim$30 nm inhomogeneous distribution of the wavelength of the emitter. By introducing the filters directly in the chip, a fully-integrated quantum photonic circuit with multiple single-photon sources and single-photon detectors can be built.

The compatibility with self-assembled QD requires cryogenic operation, wide-band tunability, and large free-spectral range (FSR) covering the inhomogenous spectral distribution of the excitonic transitions. To date, attempts at integrating a tunable filter with QD single-photon sources have only been performed in hybrid platforms, where III-V semiconductors are integrated on silicon nitride circuits\cite{elshaari2017chip,elshaari2018strain}.
In these works, filtering has been achieved using microring resonators which are either tuned by temperature or strain. Both methods suffer from limited tunability ($\sim$1 nm) and cross-talk with the emitters, which is an obstacle towards scaling up to multiple QDs emitting at the same wavelength.

\begin{figure*}
\centering
\includegraphics{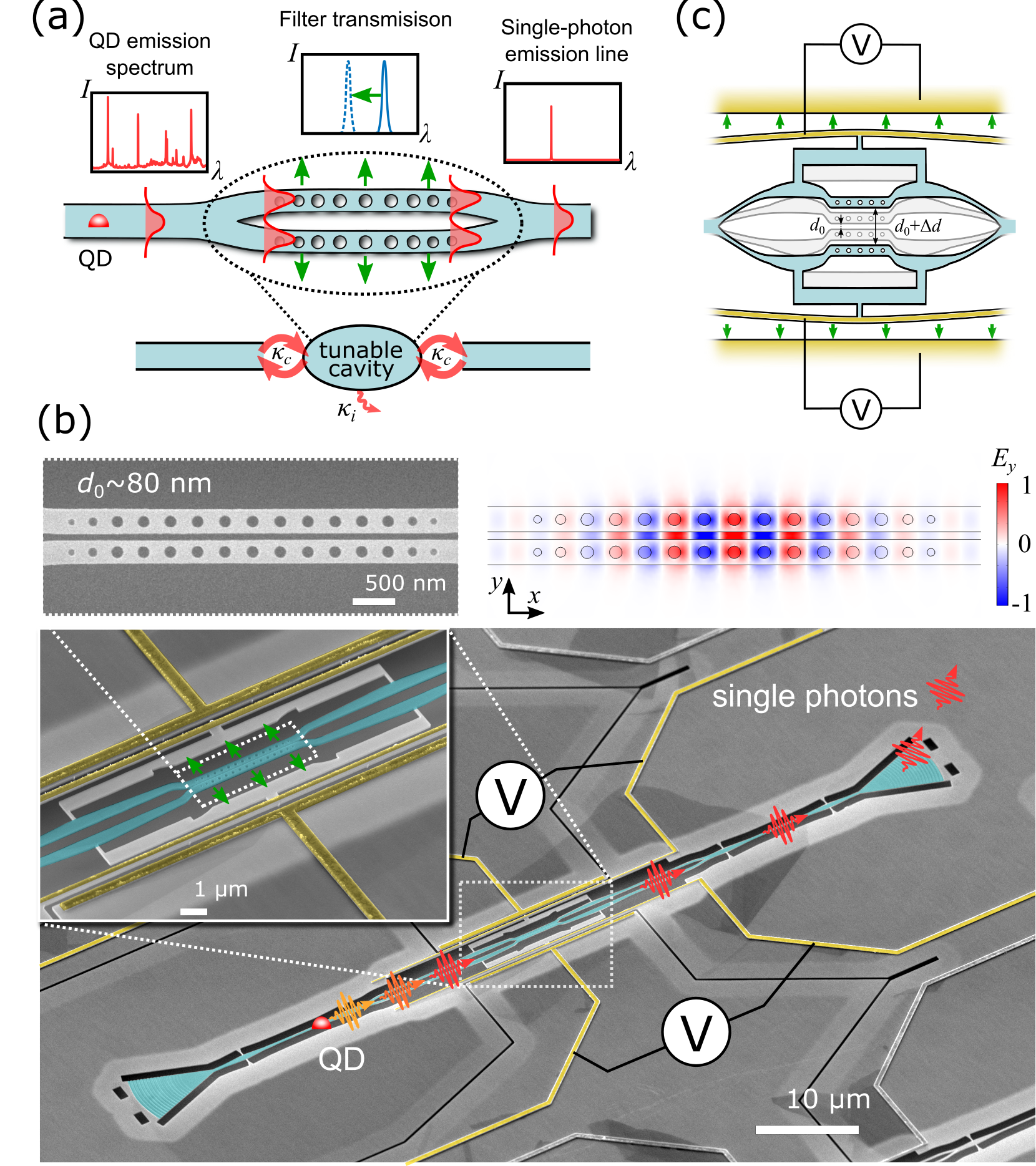}
\caption{\label{fig:concept} Reconfigurable nanomechanical single-photon filter. (a) Schematic of the working principle: the tunable filter is based on a coupled nanobeam photonic crystal cavity whose resonant wavelength is controlled mechanically, allowing the selection of one single quantum dot (QD) emission line. Red envelopes represent the optical mode distribution. $\kappa_{c}$ and $\kappa_i$ are the cavity-waveguide coupling and cavity intrinsic loss rate, respectively.
%The filter performance in direct coupling configuration is determined by the cavity-waveguide couping rate ($\kappa_{c}$) and the cavity intrinsic loss rate ($\kappa_{l}$).%
(b) False-color scanning electron micrograph (SEM) of the tunable filter with an embedded QD single-photon source (approximate position indicated by the red dot). The optical path and the electrodes are highlighted in blue and yellow, respectively. The device spectrally filters single photons (red wavepackets) emitted by the QD and rejects other emission lines (orange wavepackets). The zoom-in view shows the detail of the filter section. The direction of the electro-mechanical motion is indicated by the green arrows. The top panel shows an SEM of the coupled cavity (left) %with a narrow gap distance $d\sim80$ nm %
and its optical fundamental cavity mode from simulation (right). %, which has a strong field enhancement in the narrow gap and is well-coupled to the propagating waveguide mode.% 
(c) Schematic of the electromechanical deformation of the device. $d_0$ is the distance at zero bias and $\Delta d$ is the voltage-induced displacement.}
\end{figure*}

In this work we demonstrate for the first time a compact and tunable resonant filter in GaAs, monolithically integrated with QDs embedded in waveguides, and demonstrate filtering of single photons under non-resonant excitation. We use a nano-opto-electro-mechanical system (NOEMS)\cite{midolo2018nano} to tune the working wavelength of the filter. Thanks to the enhanced coupling between mechanical motion and light at the nano-scale, the NOEMS filter provides large wavelength tunability ($>10$ nm within 15 V) with a very small device footprint ($\sim 7\times$45 $\mu$m$^2$). Our approach is compatible with cryogenic temperature and does not interfere with the emitter itself, offering pathways for on-chip quantum experiments with multiple sources or de-multiplexing circuits.

\textbf{Device design and calibration.} Figure~\ref{fig:concept}(a) shows schematically the working principle of our reconfigurable filter, which is based on a coupled nanobeam photonic crystal cavity. Changing the distance between the nanobeams alters the coupling strength between them, resulting in a shift in the resonant wavelength of the cavity\cite{haus1987coupled}. These structures have been widely studied in the context of optomechanics in various material platforms\cite{deotare2009coupled,perahia2010electrostatically,frank2010programmable,midolo2012electromechanical,tian2019all,mouradian2017tunable}. Deotare et al. have used the coupled nanobeam photonic crystal cavity as a filter on silicon-on-insulator (SOI) and demonstrated spectral tuning with radiation pressure \cite{deotare2012all}. Chew et al. reported micro-electromechanical devices on SOI to tune the resonance of a single nanobeam photonic crystal cavity\cite{chew2010plane,chew2011nanomechanically}. Yet, none of these works have shown waveguide-integration with a quantum emitter and on-chip filtering of a quantum dot single-photon source, which is what we demonstrate here. 

The coupled cavity supports two sets of optical modes whose optical field profiles are either symmetric or anti-symmetric with respect to the nanobeam orientation\cite{haus1987coupled}. We designed a circuit that couples, by means of adiabatic Y-splitters, a single-mode waveguide to the symmetric cavity modes (cf. the mode profile shown in the top panel of Fig. \ref{fig:concept}(b)). By suppressing the coupling to anti-symmetric modes, the filter provides a maximum transmission efficiency only limited by the scattering loss in the cavity.

The fabricated device is shown in the scanning electron micrograph (SEM) in Fig.~\ref{fig:concept}(b), where InAs QDs and the tunable filter are monolithically integrated on a 160-nm thick GaAs membrane (See Supporting Information S1). The gap distance ($d$) between the two coupled nanobeams is controlled by a pair of electrostatic actuators\cite{papon2019nanomechanical}. By changing the voltage on the actuators, the filter can be tuned in resonance with a QD, allowing single photons to transit and reach a high-efficiency grating coupler\cite{zhou2018high} on the other side. Details of the filter are shown in the inset of Fig.~\ref{fig:concept}(b). A Y-shaped supporting frame is used to connect the metal electrodes to the waveguides, avoiding direct contact with the cavity region and enabling parallel motion of the nanobeams, as shown in Fig.~\ref{fig:concept}(c).
The top panel of Fig.~\ref{fig:concept}(b) shows the detail of the cavity, which is designed using the envelope equation approach outlined in Supporting Information S2.1: the lattice constant is stretched following a nearly-quadratic law to pull a single optical mode from the air-band of the photonic crystal into the bandgap\cite{Quan2011deterministic}. Mode matching between the photonic crystal Bloch modes and the waveguide modes is optimized by tapering the radii of two external holes \cite{sauvan2005modal}. %Further details on the cavity design are given in the Supplementary Information.
%Simulation of the fundamental cavity mode shows that the mode is predominantly located in the cavity holes, i.e., the air-band mode, which is deliberately chosen for achieving simultaneously a large FSR and a high $Q$ factor.
%With a gap distance ($d$) between two nanobeams as small as 80 nm, cavity mode field is greatly enhanced in the gap region, indicating strong interaction between the individual nanobeam cavities.

Our designed filter has a FSR of 25 nm and an intrinsic quality factor $Q_{i}>5.76\times10^5$. For a cavity coupled to a waveguide, the transmission coefficient is given by\cite{joannopoulos2008molding}
 \begin{equation} 
     T=\frac{\kappa^2_c}{(\kappa_c+\kappa_i)^2}=\frac{(Q_i-Q)^2}{Q_i^2},
     \label{eq1:T_Q}
 \end{equation}
where $\kappa_c$ and $\kappa_i$ are cavity-waveguide coupling rate and cavity intrinsic loss rate, respectively, $Q_c$ and $Q_i$ are the quality factors corresponding to $\kappa_c$ and $\kappa_i$, and $Q$ is the quality factor of the waveguide-loaded cavity calculated as $1/Q = 1/Q_c + 1/Q_i$. We design the cavity to support a fundamental mode with $Q\approx Q_c=6000$, which provides theoretically near-unity transmission efficiency ($\sim 98\%$). By chosing an initial nanobeam separation $d_0=80$ nm, a total wavelength shift of $\Delta\lambda = 10$ nm is expected, based on finite-element method simulations (see Supporting Information S2 for details on the device design and simulations).

\begin{figure*}
\centering
\includegraphics{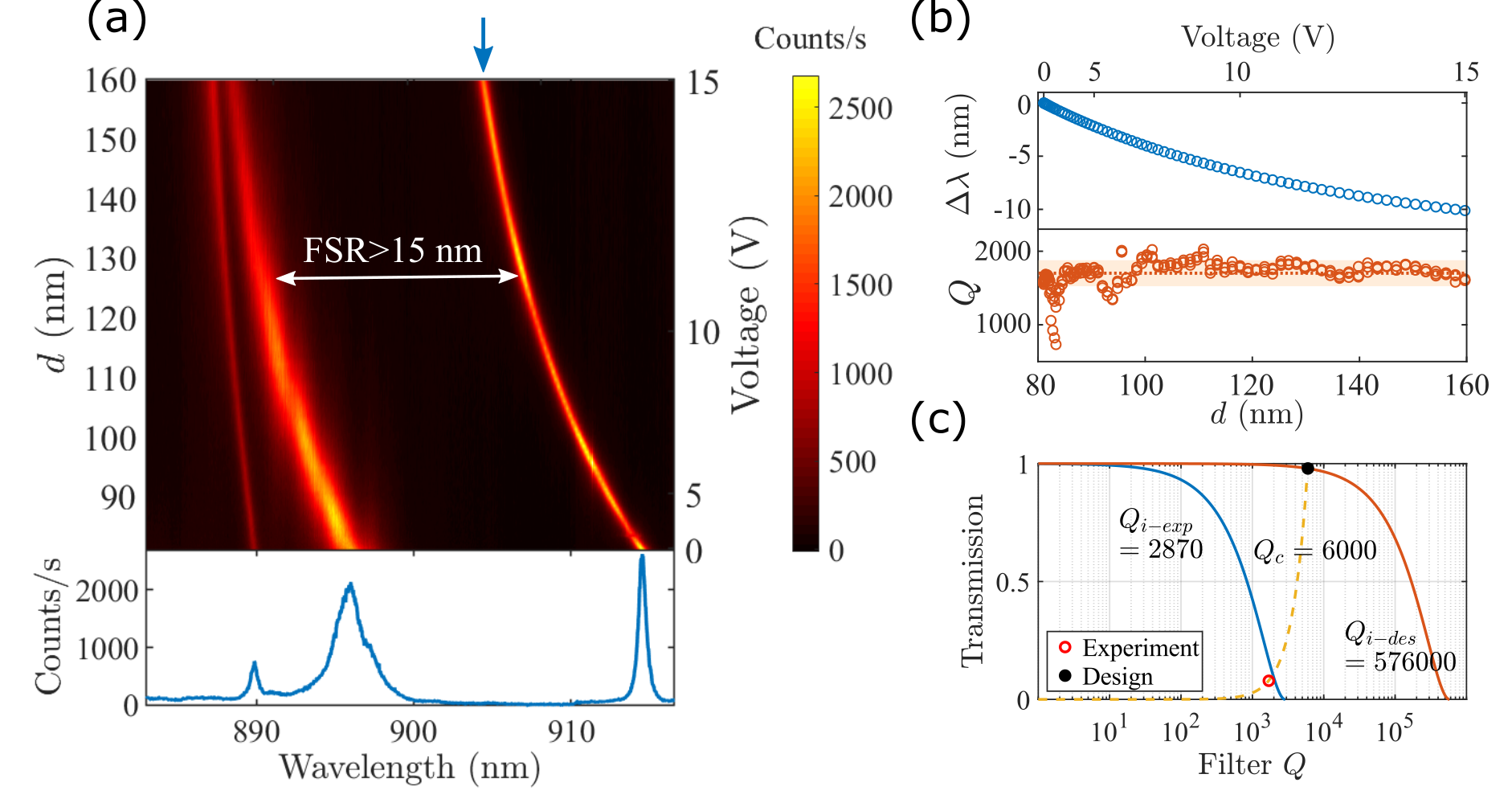}
\caption{\label{fig:cavity} Calibration of the tunable filter. (a) Transmission spectra as a function of gap distance ($d$) or voltage showing tuning of the resonant wavelengths. Transmission spectrum at 0 V is shown in the bottom panel. The fundamental cavity mode (indicated by an arrow) is red-shifted from the second-order mode by a free-spectral-range (FSR) of $>$15 nm. (b) Wavelength shift ($\Delta \lambda$) and $Q$ as a function of $d$, extracted from a Lorentzian fit of the fundamental mode in (a). (c) Trade-off between transmission efficiency and filter $Q$ for the cavity, considering intrinsic $Q$ factors from design ($Q_{i-des}$, red line) and experiment ($Q_{i-exp}$, blue line). The fabricated device has lower transmission and $Q$ (red circle) compared with design values (black dot), but retains the designed waveguide-cavity coupling, i.e., $Q_c=6000$, as indicated by the yellow dashed line.}
\end{figure*}

The NOEMS filter is characterized in a helium-flow cryostat at the temperature of 10 K. The excitation laser beam (at 800 nm wavelength) is focused on one grating coupler with high pumping power far above the saturation level of individual excitonic transitions. The photoluminescence from the QD ensemble is used as a broadband light source to probe the filter response. Light transmitted through the filter is collected from a second grating coupler and sent to a spectrometer. Figure~\ref{fig:cavity}(a) shows the collected spectra as a function of the applied voltage and mechanical displacement (see Supporting Information S2 for details on the calibration of the voltage-displacement curve). Two transmission peaks at 914.5 nm and 896 nm (cf. bottom panel of Fig. 2(a)), corresponding to the fundamental and second-order modes of the cavity, respectively, are blue-shifted by over 10 nm with 15 V applied bias. Another transmission peak around 890 nm, which is not predicted by theory, is also visible, but with a much smaller tuning range. We attribute this mode to fabrication disorder that causes localization of light into one of the two nanobeams\cite{garcia2017two} and thus is only weakly affected by the mechanical displacement. The absence of any red-shifting modes indicates that the device is indeed symmetric (See Supporting Information S2.1). The FSR of the filter extends over 15 nm and covers a large part of the QD distribution.
The transmission peaks of the fundamental mode are fitted with a Lorentzian function, and the resulting wavelength shift $\Delta \lambda$ and quality factor $Q$ are plotted in Fig.~\ref{fig:cavity}(b). The cavity resonance blue-shifts when increasing $d$, with a maximum optomechanical coupling rate $G=d\omega/dx=(2\pi)\cdot$97 GHz/nm (or $d\lambda/dx = -0.27$ nm/nm) at $d\sim80$ nm. %Within 40 nm displacement (15 V bias) for each nanobeam cavity arm, the resonance wavelength can be tuned by -10.1 nm for the current device.
The filter linewidth is 0.54 nm, corresponding to a cavity $Q\sim1700$, which is not significantly affected by the mechanical tuning. A maximum tuning range of $\Delta\lambda=-10.1$ nm is observed by increasing the bias up to 15 V, which is close to the electrostatic pull-in failure of the device. An even larger tuning range ($\Delta \lambda =-14.6$ nm) is observed on another device, albeit with a slightly lower $Q \sim 1400$ (See Supporting Information S3.3). 

By normalizing the transmission signal to that of an identical structure without photonic crystal cavity, we extract an averaged peak transmission efficiency of $8\pm1\%$. Both experimental transmission and $Q$ are lower than the theoretical values, as shown in Fig.~\ref{fig:cavity}(c). We attribute this mismatch to fabrication imperfections, e.g., resist residues, rough waveguide sidewalls, and non-uniformity of photonic-crystal holes, which cause a large drop in the intrinsic quality factor. This is verified by measuring the quality factor of cavities decoupled from the waveguides, providing an experimental $Q_{i-exp} = (2.9\pm0.3) \times 10^{3}$ (See Supporting Information S3.2). Figure~\ref{fig:cavity}(c) shows the expected transmission (cf. Eq.~\ref{eq1:T_Q}) as a function of the total filter $Q$ in the case where the intrinsic quality factor is the theoretical $Q_{i-des}$ (red line) or the experimental $Q_{i-exp}$ (blue line). The observed transmission (red dot) fits very well to the theoretical curve, indicating that transmission can be further improved by optimizing the fabrication process to reduce out-of-plane losses.

\begin{figure*}
\centering
\includegraphics{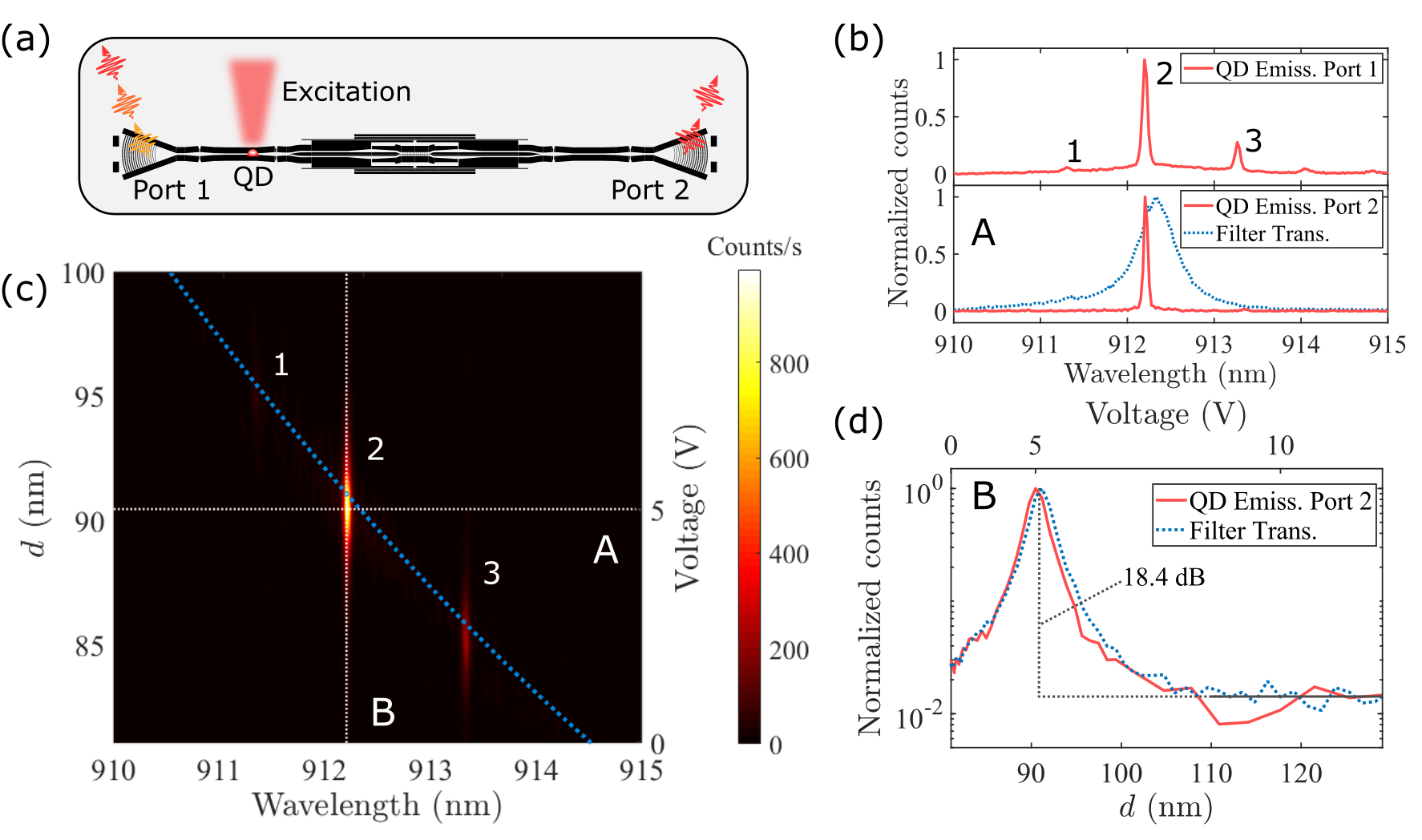}
\caption{\label{fig:QD} Selective filtering of a single quantum-dot (QD) emission line. (a) Experimental excitation and collection scheme: the QDs in the nanobeam waveguide are excited from the top, and the unfiltered and filtered photons are scattered out from grating Ports 1 and 2, respectively. (b) Top: unfiltered QD emission spectrum showing multiple lines labelled as 1, 2, and 3. Bottom: filtered QD emission with 5 V applied bias (cut line A in (c)). The blue dashed line shows the normalized transmission at the same voltage. (c) QD emission spectra collected from Port 2 as a function of the gap distance ($d$) and voltage, showing that the emission lines 1, 2, and 3 can be selectively filtered out. The blue dashed line shows the filter resonant wavelength. 
%as a guide to the eye (*** what do you mean guide to the eye? are they measurements, i.e. have you tracked the line? ***). 
(d) Filtered QD emission as a function of $d$ (cut line B in (c)) while the cavity is tuned across the QD emission, showing a maximum extinction ratio of 18.4 dB. The filter transmission from Fig.~2(a) at the QD wavelength of 912.2 nm (blue dashed line) is also shown for comparison.}
\end{figure*}

\textbf{Single-photon filtering.} Figure~\ref{fig:QD} demonstrates single-photon filtering with the electromechanically reconfigurable on-chip filter. The excitation and collection scheme is shown in Fig.~\ref{fig:QD}(a), where we use the grating named Port 1 as reference for unfiltered QD emission, and Port 2 for the filtered signal. The measured QD emission spectra from both ports are shown in Fig.~\ref{fig:QD}(b) when the cavity is tuned in resonance with the emitter. Three QD emission lines are seen in the unfiltered spectrum in the top panel. By applying a voltage bias at 5 V, we align the cavity resonance to a bright QD emission line (line 2 in the figure) at 912.2 nm, and filter out the other two lines (1 and 3), which are no longer visible in the Port 2 spectrum (cf. upper and lower panel of Fig. 3(b)). The tunable filter provides around 10 dB rejection to nearby emission lines on both sides. The advantage of a reconfigurable filter is that, in fact, one can perform selective filtering of any QD emission lines within the tuning range, as shown in the map of Fig.~\ref{fig:QD}(c). At voltage biases of 3.2 V and 6 V, the filter resonance can be aligned to two more QD lines (1 and 3, at 913.3 nm and 911.3 nm, respectively). Unlike the traditional thermo-optic tuning methods, the mechanical approach does not introduce any cross-talk with the QDs, which is confirmed by the absence of shift in the emitter energy. 

The extinction ratio (ER) is obtained from both the collected QD emission counts around 912.2 nm (cut-line B in Fig.~\ref{fig:QD}(c)) and the cavity transmission at the same wavelength, as shown in Fig.~\ref{fig:QD}(d). Comparing the peak transmission with averaged counts at gap distance $d>110$ nm ($\Delta d>30$ nm), we obtain consistent ER values of 18.5 dB and 18.3 dB from the QD and filter transmission data, respectively. The experimental ER is likely limited by fabrication disorder in the photonic crystal structure.

\begin{figure}
\centering
\includegraphics{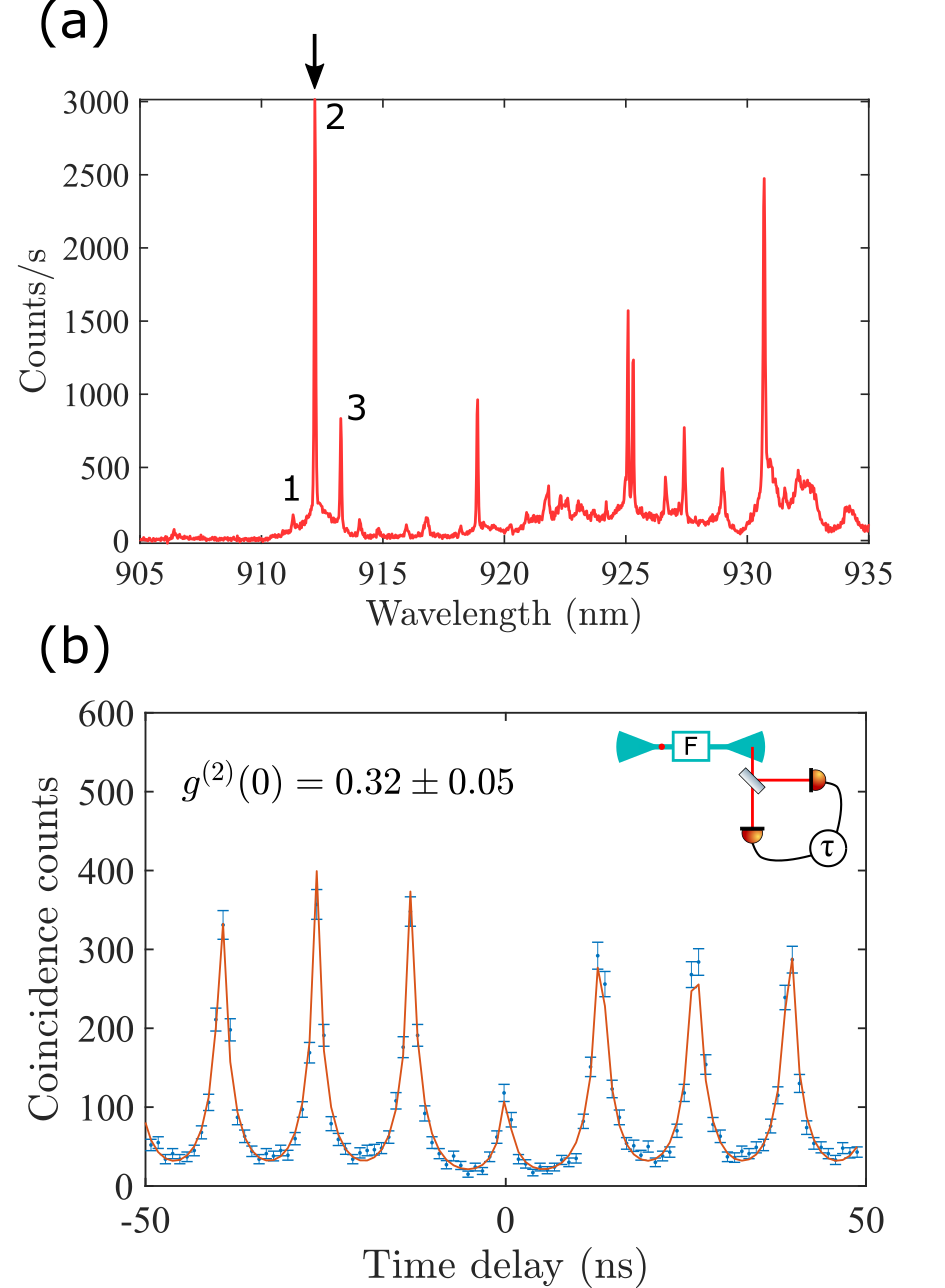}
\caption{\label{fig:g2} (a) Photoluminescence spectrum showing multiple transition lines originating from different QDs under non-resonant excitation. (b) Second-order correlation measurement of the filtered QD emission line 2 (indicated by an arrow in (a)) with the on-chip nanomechanical filter only, showing a clear anti-bunching at zero-delay.}
\end{figure}

To verify on-chip single-photon filtering, we measure the purity of the filtered single photons with a pulsed Hanbury Brown-Twiss (HBT) experiment. Figure~\ref{fig:g2}(a) shows the unfiltered emission spectrum collected from Port 1 (the same data as Fig.~\ref{fig:QD}(b) upper panel but with a broader spectral range). Multiple excitonic transmissions are visible over a $>$25 nm wavelength range due to the above-band excitation. We consider line 2 at 912.2 nm (cf. Fig. 4a) and adjust the pump to $P=0.76P_{sat}$, where $P_{sat} = 26.4$ nW is the saturation power of the transition. The measured autocorrelation of the filtered QD emission line 2 is plotted in Fig.~\ref{fig:g2}(b), where a clear signature of single-photon emission is observed at zero time delay, with $g^2(0)=0.32\pm0.05$. We note that the QD emission collected from Port 2 after the on-chip filter is sent directly to the HBT set-up without additional filtering, as schematically shown in the inset. The $g^2(0)$ is still limited by the above-band excitation scheme, the high QD density ($>$10 $\mu$m$^{-2}$), and the finite ER of the on-chip filter. Yet, our recorded $g^2(0)$ is only slightly larger than the typical $g^2(0)=0.1-0.2$ obtained with off-chip filtering using above-band excitation\cite{daveau2017efficient,elshaari2017chip,elshaari2018strain,papon2019nanomechanical,schnauber2019deterministic}, indicating comparable single-photon filtering performance.

\textbf{Discussion.} Transmission (or insertion loss) of the on-chip filter is a very important figure-of-merit for quantum experiments, and the current $\sim 8\%$ transmission of our device can be improved through different methods. Equation~\ref{eq1:T_Q} shows that there is a trade-off between transmission and $Q$ (or filter bandwidth) for a given $Q_i$, as shown in Fig.~\ref{fig:cavity}(c) for both $Q_{i-des}$ and $Q_{i-exp}$. The transmission of the fabricated device (the red circle) is lower than the design (the black dot), but it fits very well to the theoretical transmission calculated by fixing $Q_c$ to the design value (yellow dashed line), indicating nearly no loss from the cavity-waveguide interface. Higher transmission can be achieved by lowering $Q_c$ (e.g., reducing the number of holes forming the cavity), at the cost of a lower total $Q$. Alternatively, the intrinsic quality factor $Q_i$ can be increased by optimizing the fabrication process, which results in simutaneoulsy improved transmission and narrower filter linewidth. Using a hard mask for reactive ion etching of GaAs\cite{gonzalez2014fabrication} or performing surface passivation\cite{guha2017surface,liu2018single} could greatly improve the intrinsic cavity $Q$. As a quantitative prospect, nanobeam photonic crystal cavities with $Q = 2\times10^4$ have been demonstrated in suspended GaAs membrane platform\cite{enderlin2012high}, suggesting that $T>60$\% is within reach with the proposed design. Increasing the $Q$ would also enable applications such as filtering of phonon sidebands, that requires around 40 GHz bandwidth ($Q\sim 10^4$) to produce high-indistinguishability photons\cite{lodahl2015interfacing}.

\textbf{Conclusion \& Outlook.} We have demonstrated single-photon filtering with a nanomechanically reconfigurable nanophotonic cavity, which is monolithically integrated with a QD single-photon source. The device is operated at $<$10 K without any cross-talk with the emitter. Thanks to the enhanced opto-electro-mechanical interaction at the nanoscale, a large tunablity of $>10$ nm at a bias of 15 V can be achieved, comparable to the inhomogeneous spectral distribution of QD emission. The filter design, together with the small footprint of $7\times45$ $\mu$m$^2$, is intrinsically low-loss and scalable. A high ER of $> 18$ dB and a large FSR $> 15$ nm ensure good single-photon purity.

The device reported here has the salient features of a broadband device which could be replicated for multiple emitters on the same chip. Combined with a $p$-$i$-$n$ diode structure for the Stark tuning of the excitonic transitions\cite{kirvsanske2017indistinguishable}, on-chip sources can be realized and scaled by controlling both the emitter and the filter resonances\cite{petruzzella2018quantum}. Furthermore, positioning the QDs inside the cavity and tuning the cavity in-resonance with the QD emission, promises to improve both single-photon generation efficiency and indistinguishability with Purcell effect\cite{midolo2012spontaneous,iles2017phonon}. A single on-chip filter can also replace multiple off-chip filtering setups, which are required in multi-path experiments, e.g. on-chip demultiplexing\cite{papon2019nanomechanical,lenzini2017active}, where the single-photon source is routed to many different outputs. The tunable filter reported in this work constitutes an important building block for scaling these experiments further, towards a fully-integrated photonic quantum information processing with multiple qubits.

\begin{acknowledgement}

We gratefully acknowledge financial support from Innovationsfonden (Quantum Innovation Center
QUBIZ), Danmarks Grundforskningsfond
(DNRF) (Center for Hybrid Quantum Networks (Hy-Q)),
H2020 European Research Council (ERC) (SCALE) \& ERC-Proof of Concept (FIPS 790206), Bundesministerium für Bildung und Forschung (BMBF)
(16KIS0867, Q.Link.X), Deutsche Forschungsgemeinschaft
(DFG) (TRR 160), Styrelsen for Forskning og Innovation (FI)
(5072-00016B QUANTECH).

The authors thank Henri Thyrrestrup for helpful discussion.

\end{acknowledgement}

\begin{suppinfo}

\end{suppinfo}

\bibliography{Main.bib}

\end{document}